\documentstyle[12pt,emlines2]{article}
\newcommand{\NP}[1]{Nucl. \ Phys.}
\newcommand{\PL}[1]{Phys. \ Lett.}
\newcommand{\p}[1]{\partial}
\newcommand{\PRL}[1]{Phys.\ Rev.\ Lett. }
\newcommand{\MPL}[1] { Mod. Phys. Lett. }
\newcommand{\IJMP}[1] { Int. J. Mod. Phys. }
\begin{document}

\title{
\begin{flushright}
{\small SMI/TH-06-96 }
\end{flushright}
\vspace{10mm}
Parquet Approximation in Large $N$  Matrix Theories}
\author{
I.Ya.Aref'eva, \thanks{Steklov Mathematical Institute,
Vavilov 42, GSP-1, 117966, Moscow,
e-mail: arefeva@class.mian.su} \\
\\ and\\ A.P.Zubarev \thanks {Steklov Mathematical Institute,
Vavilov 42, GSP-1, 117966, Moscow, e-mail: zubarev@class.mian.su}}
\date {$~$}
\maketitle

\begin {abstract}
A parquet approximation (generalized ladder diagrams)
in matrix models is considered.
By means of numerical calculations we demonstrate that in the large $N$
limit the parquet approximation gives an excellent
agreement with exact results.
\end {abstract}


\section{Introduction}

The parquet approximation (or generalized ladder approximation)
was introduced by Landau, Abrikosov and Khalatnikov in their
famous consideration of the high energy behavior in quantum
electrodynamics \cite{LAK}. Later on  this approximation has been used
for various models of quantum field theory \cite{TM}.
The results of the parquet approximation
for weak coupling are in  agreement with   the renormalization group
approach \cite{BSh}.  Let us recall that the original aim of Landau
at all  was to develop a non-perturbative method in quantum field theory.
The parquet approximation leads to a closed system of integral equations
which have meaning not only for small but also for the large coupling constant.
In this sense the method of the parquet approximation is a
non-perturbative one.
There are  arguments \cite{LAK,TM} that this approximation is able to catch
the asymptotic behavior in quantum field theory at small distances even
in the strong coupling regime.  However, there is a serious problem here,
see \cite{BSh}, because it is very difficult to estimate an error
of this approximation  in realistic models for  strong coupling, since
we have few information about the true behaviour of the model in this regime.
One can control the strong coupling regime only for some simple models such as
vector models \cite{AA} and  low dimensional matrix
models \cite{BIPZ}  in the large $N$ limit.

The aim of this letter is to investigate the parquet approximation for
matrix models in the large $N$ limit.
By using numerical calculations we demonstrate that the parquet
approximation proves to be a  surprisingly good approximation for $N\times
N$ matrix models in the large $N$ limit. To this end we compare the numbers
of parquet diagrams with  suitable weights with the total number of diagrams
with the same weights in the large $N$ limit.  As the weight of diagrams we
take the corresponding powers of $N^{-1}$ and  coupling constants.  It
is occurred that in the large $N$ limit these two numbers that are functions
on coupling constants and the number of external lines almost coincide
for all values of the coupling constants including the strong coupling
limit.

The behavior of the weighted number of all diagrams in the large $N$ limit
is well known \cite{BIPZ}. It is given by the large $N$ limit of
the corresponding zero-dimensional matrix model. We find the weighted
numbers of  parquet diagrams for large
$N$ matrix theories with  the quartic and
cubic interactions.  We find these numbers as solutions of
closed sets of equations for parquet correlation functions in the leading
order of $N^{-1}$ expansion.
These equations are  simpler  than
the  parquet equations in \cite{TM} since in our
equations $u$-channel terms do not contribute.

As it is known the large $N$ limit in QCD enables us  to understand
qualitatively certain striking phenomenological features of strong
interactions \cite{tH}-\cite{AS}.  To perform  analytical
investigations one needs to compute the sum of all planar diagrams.
At present this problem has been solved only in a few simple models
such as zero- and one-dimensional matrix models and two-dimensional QCD.
It was suggested \cite{Witten}
that in the large $N$ limit the dynamics is dominated by a
master field. The problem of finding of the master
field has been discussed in many works \cite {Haan}-\cite {Gopak}.
Recently the
problem of construction of the master field
has been solved in \cite{AV}.
It was shown that the master field satisfies to standard equations of
relativistic quantum field theory but fields are quantized according to a
new rule.  An explicit solution of these equations is a rather
non-trivial problem and it seems reasonable to develop some approximated
schemes to study the master field.

There were several attempts
of an approximated treatment of the planar theory \cite {Sl}-\cite{Fer}.
In particular, there were  approximations dealing with
counting only some part of planar
diagrams.  In \cite{AAV,AZ}  a so-called  half-planar
approximation to the planar theory was considered.
 The distinguished feature of this
approximation is that one can write down a closed set of Schwinger-Dyson-like
equations for a finite number of Green's functions.
The half-planar approximation gives
a good agreement with the known exact answers for the large $N$ models
 for a rather wide range of coupling constant \cite{AZ}.
However, this approximation does not reproduce the correct
strong coupling limit
of exact planar Green's functions. A recursive scheme which could improve
the half-planar approximation
was proposed in \cite{Ar96}. It turns out that this scheme is very closed
to the approximation adopted in this paper.

The paper is organized as follows.
In Section 2 we give a general definition for
the planar parquet approximation.
In Sections 3 and 4 we compare   the planar parquet and
the planar solutions
of zero-dimensional matrix models.
We find  a surprising coincidence of these solutions
for all range of the coupling
constant up to 0.1 percent.  Moreover, we find that the planar parquet
solutions for the quartic and cubic interactions have  phase transition
points which  also exist in the planar solutions of the corresponding
one-matrix models. The values of phase transition points for the planar
parquet and  the planar approximations are
turned out to be very closed.

\section{ Planar Parquet Equations}

In this section we define the planar parquet approximation
for the $d$-dimensional  matrix model with
the  cubic and quartic interaction terms,
\begin{equation}
\label{Action}
S= \int d^dx Tr [ \frac{1}{2}
\partial M \cdot \partial M +
\frac{1}{2} m^2 M^2
-\frac{\lambda}{3 N^{1/2}} M^3 +
\frac{g}{4 N} M^4].
\end{equation}
Here $M=(M_{ij}(x)),~$ $i,j=1,...,N~$ is an Hermitian matrix field.
The planar Green's functions   are defined as $N\to \infty$ limit
of invariant correlation functions of products of matrices
\begin{equation}
\label{Greens}
\Pi_n(x_1,...,x_n)=\lim_{N \to
\infty}\frac{1}{N^{1+n/2}}\int DM Tr( M(x_1)...
M(x_n))\exp (-S)  .
\end{equation}
They   satisfy to the planar Schwinger-Dyson equations \cite{Haan,Ar81}
\begin {eqnarray}
(-\bigtriangleup  +m^{2})_{x_{1}}\Pi_{n}(x_{1},...,x_{n})-
\lambda \Pi_{n+1}(x_{1},x_{1},x_{2},...,x_{n})+
g \Pi_{n+2}(x_{1},x_{1},x_{1},x_{2},...,x_{n})-
\nonumber
\\
\sum _{m=2}^{n}\delta (x_{1}-x_{m})\Pi_{m-2}(x_{2},...,x_{m-1})
\Pi_{n-m}(x_{m+1},...,x_{n})=0.~~~~~~~~~~~~~~~~~~
\label{PSDE}
\end   {eqnarray}
These equations look almost as the Schwinger-Dyson equations for a
scalar theory but there is an essential difference in the form
of the Schwinger's terms. Planar Green's functions are
symmetric only under cyclic permutations of arguments.
We define the planar parquet  approximation
as an  approximated perturbative
solution of  equations (\ref{PSDE}) that takes into
account only a part of full series on coupling constants.
This part is
specified by a requirement that all three- and four-point vertex parts (1PI
parts) are composed from two-particle reducible (2PR) diagrams. Note that the
perturbative definition is meaningful since the perturbative solution of
the planar Schwinger-Dyson equations converges at least for asymptotically
free interactions.  So, this is  the case
for the interaction (\ref{Action}) if we average the signs of the couplings
in the suitable way as well as  for  $SU(\infty)$ QCD.

More precisely, the definition of the planar parquet approximation
can be done using a notion of a so-called skeleton expansion.
The skeleton expansion contains only a subset of all
planar diagrams. In this subset one takes only those diagrams that
contain no propagators, three- and four-vertices insertions.
Following 't Hooft \cite{tH1}
we call the two-, three- and four-point functions
the "basic Green's functions". In the four-dimensional
space-time for the cases of interaction (\ref{Action})
as well as of Yang-Mills theory just these vertex functions
contain divergences.
To get the planar parquet approximation
one has to take the set of skeleton diagrams and
replace the bare propagator and bare three- and four-vertices
by the parquet propagator and the corresponding vertex functions.

The parquet basic Green's functions are defined as  solutions of the set
of equations that is presented below. The propagator is defined
as a solution of the exact Schwinger-Dyson  equation

\vspace{5mm}

\unitlength=1.00mm
\special{em:linewidth 0.4pt}
\linethickness{0.4pt}
\begin{picture}(145.00,17.60)
\special{em:linewidth 0.4pt}
\linethickness{0.4pt}
\put(38.00,15.00){\makebox(0,0)[cc]{$=$}}
\put(60.00,15.00){\line(-1,0){12.00}}
\put(133.00,15.00){\circle*{5.20}}
\put(115.00,15.00){\line(1,0){8.00}}
\put(90.00,15.00){\circle*{5.20}}
\put(72.00,15.00){\line(1,0){8.00}}
\put(66.00,15.00){\makebox(0,0)[cc]{$+$}}
\put(108.00,15.00){\makebox(0,0)[cc]{$+$}}
\special{em:linewidth 1.2pt}
\linethickness{1.2pt}
\emline{122.00}{15.00}{1}{145.00}{15.00}{2}
\emline{122.00}{15.00}{3}{133.00}{16.67}{4}
\emline{133.00}{13.33}{5}{123.00}{15.00}{6}
\emline{30.00}{15.00}{7}{17.00}{15.00}{8}
\emline{79.00}{15.00}{9}{90.00}{16.67}{10}
\emline{90.00}{13.33}{11}{80.00}{15.00}{12}
\emline{90.00}{15.00}{13}{102.00}{15.00}{14}
\end{picture}
\vspace{-20mm}
\begin{equation}
\label{SDD}
\end{equation}
(all contributions of tadpole diagrams are dropped out).
Here

\vspace{10mm}

\unitlength=1mm
\begin{picture}(112.00,15.00)
\special{em:linewidth 0.4pt}
\linethickness{0.4pt}
\put(90.00,15.00){\line(1,0){15.00}}
\put(47.00,15.00){\makebox(0,0)[lc]{$=~~~D(p)~,$}}
\put(32.67,12.00){\makebox(0,0)[cc]{$_p$}}
\put(97.67,12.00){\makebox(0,0)[cc]{$_p$}}
\put(112.00,15.00){\makebox(0,0)[lc]{$=~~~\frac{1}{p^2+m^2}$}}
\special{em:linewidth 1.2pt}
\linethickness{1.2pt}
\emline{25.00}{15.00}{1}{40.00}{15.00}{2}
\end{picture}

\vspace{-20mm}

$$~$$
are the full (within the parquet approximation) and  the bare
propagators, respectively,

\vspace{5mm}

\unitlength=1.00mm
\special{em:linewidth 0.4pt}
\linethickness{0.4pt}
\begin{picture}(107.00,20.33)
\put(35.00,15.00){\circle*{5.20}}
\put(35.00,15.00){\line(-1,0){5.67}}
\put(35.00,15.00){\line(1,1){4.00}}
\put(35.00,15.00){\line(1,-1){4.00}}
\put(29.00,13.00){\makebox(0,0)[cc]{$_p$}}
\put(40.00,17.67){\makebox(0,0)[cc]{$_q$}}
\put(48.00,15.00){\makebox(0,0)[lc]{$=~~~\Gamma_3 (p,q)~,$}}
\put(95.00,15.00){\circle*{5.20}}
\put(91.00,19.00){\line(1,-1){8.00}}
\put(99.00,19.00){\line(-1,-1){8.00}}
\put(91.67,9.67){\makebox(0,0)[cc]{$_p$}}
\put(92.67,20.33){\makebox(0,0)[cc]{$_q$}}
\put(100.00,18.00){\makebox(0,0)[cc]{$_k$}}
\put(107.00,15.00){\makebox(0,0)[lc]{$=~~~\Gamma _4 (p,q,k)$}}
\end{picture}

\vspace{-20mm}

$$~$$
are three- and four-point vertices.
The three-point parquet vertex
is defined as a solution of the following  equation

\vspace{5mm}

\unitlength=1.00mm
\special{em:linewidth 0.4pt}
\linethickness{0.4pt}
\begin{picture}(142.00,21.67)
\special{em:linewidth 0.4pt}
\linethickness{0.4pt}
\put(63.00,13.33){\line(0,1){3.33}}
\put(18.00,15.00){\circle*{5.20}}
\put(13.00,15.00){\line(1,0){5.00}}
\put(21.67,18.67){\line(-1,-1){3.67}}
\put(18.00,15.00){\line(1,-1){3.67}}
\put(105.00,20.00){\circle*{3.33}}
\put(105.00,10.00){\circle*{3.33}}
\put(105.00,20.00){\line(1,0){3.67}}
\put(105.00,10.00){\line(1,0){3.67}}
\put(95.00,15.00){\line(-1,0){4.00}}
\put(29.33,15.00){\makebox(0,0)[cc]{$=$}}
\put(42.00,15.00){\line(6,5){4.00}}
\put(42.00,15.00){\line(6,-5){4.00}}
\put(38.00,15.00){\line(1,0){4.00}}
\put(63.00,15.00){\circle*{5.20}}
\put(63.00,15.00){\line(-1,0){4.67}}
\put(75.00,19.00){\line(1,0){4.00}}
\put(75.00,11.00){\line(1,0){4.00}}
\put(73.00,9.00){\rule{4.00\unitlength}{12.00\unitlength}}
\put(125.00,15.00){\circle*{5.20}}
\put(125.00,15.00){\line(-1,0){4.67}}
\put(10.33,15.00){\makebox(0,0)[cc]{$_p$}}
\put(21.00,21.00){\makebox(0,0)[cc]{$_q$}}
\put(20.33,10.33){\makebox(0,0)[cc]{$_{-p-q}$}}
\put(66.67,21.00){\makebox(0,0)[cc]{$_l$}}
\put(66.67,11.00){\makebox(0,0)[cc]{$_{l-p}$}}
\put(81.33,20.33){\makebox(0,0)[cc]{$_q$}}
\put(88.00,17.00){\makebox(0,0)[cc]{$_p$}}
\put(83.67,15.00){\makebox(0,0)[cc]{$+$}}
\put(111.00,21.67){\makebox(0,0)[cc]{$_q$}}
\put(113.00,15.00){\makebox(0,0)[cc]{$+$}}
\put(117.33,16.67){\makebox(0,0)[cc]{$_p$}}
\put(130.00,19.33){\makebox(0,0)[cc]{$_l$}}
\put(142.00,18.00){\makebox(0,0)[cc]{$_q$}}
\put(130.00,11.33){\makebox(0,0)[cc]{$_{l-p}$}}
\put(140.00,17.00){\line(-2,-1){4.00}}
\put(136.00,15.00){\line(2,-1){4.00}}
\special{em:linewidth 1.2pt}
\linethickness{1.2pt}
\emline{125.00}{16.67}{1}{136.00}{15.00}{2}
\emline{136.00}{15.00}{3}{125.00}{13.33}{4}
\emline{95.00}{15.00}{5}{105.00}{20.00}{6}
\emline{105.00}{20.00}{7}{105.00}{10.00}{8}
\emline{105.00}{10.00}{9}{95.00}{15.00}{10}
\emline{75.00}{19.00}{11}{63.00}{16.67}{12}
\emline{63.00}{16.67}{13}{63.00}{13.33}{14}
\emline{63.00}{13.33}{15}{75.00}{11.00}{16}
\put(50.00,15.00){\makebox(0,0)[cc]{$+$}}
\put(57.00,17.00){\makebox(0,0)[cc]{$_p$}}
\put(96.00,15.00){\circle*{3.33}}
\put(98.67,19.33){\makebox(0,0)[cc]{$_l$}}
\put(99.00,11.00){\makebox(0,0)[cc]{$_{l-p}$}}
\end{picture}
\vspace{-20mm}
\begin{equation}
\label{MG3}
\end{equation}
Equation (\ref{MG3}) means that any 1PI 3-point
diagram (except the bare one) can be decomposed into
three- and four-point subdiagrams connected by two internal lines
so that the four point subdiagram is 2PI with respect to those two lines
that connect  the four- and three-point subdiagrams.
It is evident that such four-point subdiagram can be
either  the bare vertex either  a "dumb-bells" diagrams or
a $V$-vertex.
Here the filled vertical   box $V$  (horizontal box $H$)

\vspace{5mm}

\unitlength=1mm
\special{em:linewidth 0.4pt}
\linethickness{0.4pt}
\begin{picture}(110.00,22.00)
\put(30.00,13.00){\rule{12.00\unitlength}{4.00\unitlength}}
\put(95.00,9.00){\rule{4.00\unitlength}{12.00\unitlength}}
\put(32.00,11.00){\line(0,1){8.00}}
\put(40.00,19.00){\line(0,-1){8.00}}
\put(93.00,19.00){\line(1,0){8.00}}
\put(93.00,11.00){\line(1,0){8.00}}
\put(32.00,9.00){\makebox(0,0)[cc]{$_p$}}
\put(32.00,22.00){\makebox(0,0)[cc]{$_q$}}
\put(40.00,22.00){\makebox(0,0)[cc]{$_k$}}
\put(91.00,11.00){\makebox(0,0)[cc]{$_p$}}
\put(91.00,19.00){\makebox(0,0)[cc]{$_q$}}
\put(103.00,19.00){\makebox(0,0)[cc]{$_k$}}
\put(50.00,15.00){\makebox(0,0)[lc]{$=~~~H(p,q,k)~,$}}
\put(110.00,15.00){\makebox(0,0)[lc]{$=~~~V(p,q,k)~$}}
\end{picture}

\vspace{-20mm}

$$~$$
is  the part of the four-point
vertex function that is 2PR in the $t$-channel
($s$-channel) and is not 2PR in the $s$-channel ($t$-channel).
The two particle reducibility in the $s$-channel
means that the $H$-vertex can be represented
in the following  way

\vspace{5mm}

\unitlength=1.00mm
\special{em:linewidth 0.4pt}
\linethickness{0.4pt}
\begin{picture}(127.00,46.67)
\special{em:linewidth 0.4pt}
\linethickness{0.4pt}
\put(48.00,15.33){\line(0,1){3.33}}
\put(117.00,17.00){\line(0,1){1.67}}
\put(117.00,15.00){\line(0,1){2.00}}
\put(48.00,17.00){\line(0,-1){1.67}}
\put(25.00,39.00){\rule{12.00\unitlength}{4.00\unitlength}}
\put(34.00,11.00){\rule{4.00\unitlength}{12.00\unitlength}}
\put(48.00,17.00){\circle*{5.20}}
\put(32.00,13.00){\line(1,0){3.00}}
\put(35.00,21.00){\line(-1,0){3.00}}
\put(48.00,18.67){\line(1,0){4.00}}
\put(48.00,15.33){\line(1,0){4.00}}
\put(27.00,45.00){\line(0,-1){8.00}}
\put(35.00,37.00){\line(0,1){8.00}}
\put(105.00,12.00){\circle*{3.33}}
\put(105.00,22.00){\circle*{3.33}}
\put(117.00,17.00){\circle*{5.20}}
\put(105.00,22.00){\line(-1,0){3.67}}
\put(101.33,12.00){\line(1,0){3.67}}
\put(117.00,18.67){\line(1,0){4.33}}
\put(117.00,15.33){\line(1,0){4.33}}
\put(68.00,11.00){\rule{4.00\unitlength}{12.00\unitlength}}
\put(70.00,21.00){\line(-1,0){4.00}}
\put(70.00,13.00){\line(-1,0){4.00}}
\put(83.00,13.00){\circle*{3.33}}
\put(83.00,21.00){\circle*{3.33}}
\put(45.00,41.00){\makebox(0,0)[cc]{$=$}}
\put(48.00,18.67){\line(0,-1){3.33}}
\put(75.00,41.00){\circle*{5.20}}
\put(75.00,42.67){\line(1,0){4.00}}
\put(75.00,39.33){\line(1,0){4.00}}
\put(113.00,45.00){\circle*{3.33}}
\put(113.00,37.00){\circle*{3.33}}
\put(113.00,37.00){\line(1,0){3.67}}
\put(113.00,45.00){\line(1,0){3.67}}
\put(89.00,41.00){\makebox(0,0)[cc]{$+$}}
\put(127.00,41.00){\makebox(0,0)[cc]{$+$}}
\put(26.00,35.00){\makebox(0,0)[cc]{$_p$}}
\put(25.33,46.67){\makebox(0,0)[cc]{$_q$}}
\put(36.33,46.67){\makebox(0,0)[cc]{$_k$}}
\put(36.33,35.00){\makebox(0,0)[cc]{$_{-p-q-k}$}}
\put(56.33,42.67){\makebox(0,0)[cc]{$_q$}}
\put(56.33,38.67){\makebox(0,0)[cc]{$_p$}}
\put(66.67,45.00){\makebox(0,0)[cc]{$_l$}}
\put(81.67,42.67){\makebox(0,0)[cc]{$_k$}}
\put(96.00,38.67){\makebox(0,0)[cc]{$_p$}}
\put(96.00,43.33){\makebox(0,0)[cc]{$_q$}}
\put(105.00,45.67){\makebox(0,0)[cc]{$_l$}}
\put(119.33,45.00){\makebox(0,0)[cc]{$_k$}}
\put(124.67,18.67){\makebox(0,0)[cc]{$_k$}}
\put(112.00,22.33){\makebox(0,0)[cc]{$_l$}}
\put(98.00,22.00){\makebox(0,0)[cc]{$_q$}}
\put(98.00,12.00){\makebox(0,0)[cc]{$_p$}}
\put(93.00,17.00){\makebox(0,0)[cc]{$+$}}
\put(90.00,21.00){\makebox(0,0)[cc]{$_k$}}
\put(76.67,24.00){\makebox(0,0)[cc]{$_l$}}
\put(62.00,21.00){\makebox(0,0)[cc]{$_q$}}
\put(62.00,13.00){\makebox(0,0)[cc]{$_p$}}
\put(58.33,17.00){\makebox(0,0)[cc]{$+$}}
\put(54.00,18.67){\makebox(0,0)[cc]{$_k$}}
\put(43.00,22.00){\makebox(0,0)[cc]{$_l$}}
\put(28.33,21.00){\makebox(0,0)[cc]{$_q$}}
\put(28.00,13.00){\makebox(0,0)[cc]{$_p$}}
\put(102.00,41.00){\line(-2,1){4.00}}
\put(98.00,39.00){\line(2,1){4.00}}
\put(63.00,41.00){\line(-2,1){4.00}}
\put(63.00,41.00){\line(-2,-1){4.00}}
\put(83.00,13.00){\line(1,0){4.00}}
\put(83.00,21.00){\line(1,0){4.00}}
\special{em:linewidth 1.2pt}
\linethickness{1.2pt}
\emline{36.00}{21.00}{1}{48.00}{18.67}{2}
\emline{48.00}{15.33}{3}{36.00}{13.00}{4}
\emline{70.00}{21.00}{5}{83.00}{21.00}{6}
\emline{83.00}{21.00}{7}{83.00}{13.00}{8}
\emline{83.00}{13.00}{9}{70.00}{13.00}{10}
\emline{117.00}{18.67}{11}{105.00}{22.00}{12}
\emline{105.00}{22.00}{13}{105.00}{12.00}{14}
\emline{105.00}{12.00}{15}{117.00}{15.33}{16}
\emline{102.00}{41.00}{17}{113.00}{45.00}{18}
\emline{113.00}{45.00}{19}{113.00}{37.00}{20}
\emline{113.00}{37.00}{21}{102.00}{41.00}{22}
\emline{63.00}{41.00}{23}{75.00}{42.67}{24}
\emline{75.00}{42.67}{25}{75.00}{39.33}{26}
\emline{75.00}{39.33}{27}{63.00}{41.00}{28}
\put(112.00,12.00){\makebox(0,0)[cc]{$_{p+q-l}$}}
\put(77.00,11.00){\makebox(0,0)[cc]{$_{p+q-l}$}}
\put(43.00,12.00){\makebox(0,0)[cc]{$_{p+q-l}$}}
\put(68.00,38.00){\makebox(0,0)[cc]{$_{p+q-l}$}}
\put(106.00,37.00){\makebox(0,0)[cc]{$_{p+q-l}$}}
\end{picture}
\vspace{-20mm}
\begin{equation}
\label{MH}
\end{equation}

Equation (\ref{MH})  describes the structure
of $H$-type diagrams. It means that   any $H$-diagram  can
be decomposed into two four-point subdiagrams connected by
two internal lines so that at least one of
these subdiagrams is 2PI in the $s$-channel.
The similar relation holds for $V$-vertex.
The vertices $V$ and $H$ are related  by the cyclic permutation of
external points
\begin{equation}
\label{HV}
V(p,q,k)=H(q,k,-p-q-k).
\end{equation}

The parquet approximation for the four-vertex
is  defined by

\vspace{5mm}

\unitlength=1.00mm
\special{em:linewidth 0.4pt}
\linethickness{0.4pt}
\begin{picture}(140.00,21.00)
\special{em:linewidth 0.4pt}
\linethickness{0.4pt}
\put(23.00,15.00){\circle*{5.20}}
\put(65.00,13.00){\rule{12.00\unitlength}{4.00\unitlength}}
\put(100.00,9.00){\rule{4.00\unitlength}{12.00\unitlength}}
\put(89.00,15.00){\makebox(0,0)[cc]{$+$}}
\put(106.00,19.00){\line(-1,0){8.00}}
\put(98.00,11.00){\line(1,0){8.00}}
\put(75.00,11.00){\line(0,1){8.00}}
\put(67.00,19.00){\line(0,-1){8.00}}
\put(20.00,18.00){\line(1,-1){6.00}}
\put(26.00,18.00){\line(-1,-1){6.00}}
\put(35.00,15.00){\makebox(0,0)[cc]{$=$}}
\put(125.00,11.00){\circle*{3.33}}
\put(125.00,19.00){\circle*{3.33}}
\put(133.00,19.00){\circle*{3.33}}
\put(133.00,11.00){\circle*{3.33}}
\put(133.00,11.00){\line(1,0){3.67}}
\put(133.00,19.00){\line(1,0){3.67}}
\put(125.00,19.00){\line(-1,0){3.67}}
\put(121.33,11.00){\line(1,0){3.67}}
\put(114.00,15.00){\makebox(0,0)[cc]{$+$}}
\put(44.00,12.00){\line(1,1){6.00}}
\put(44.00,18.00){\line(1,-1){6.00}}
\put(118.00,11.00){\makebox(0,0)[cc]{$_p$}}
\put(118.00,19.00){\makebox(0,0)[cc]{$_q$}}
\put(140.00,19.00){\makebox(0,0)[cc]{$_k$}}
\put(129.00,20.67){\makebox(0,0)[cc]{$_{l+q}$}}
\put(123.00,15.67){\makebox(0,0)[cc]{$_l$}}
\put(129.00,9.00){\makebox(0,0)[cc]{$_{l-p}$}}
\put(139.00,15.33){\makebox(0,0)[cc]{$_{l+q+k}$}}
\put(94.67,11.00){\makebox(0,0)[cc]{$_p$}}
\put(94.67,19.00){\makebox(0,0)[cc]{$_q$}}
\put(108.67,19.00){\makebox(0,0)[cc]{$_k$}}
\put(77.33,20.00){\makebox(0,0)[cc]{$_k$}}
\put(68.67,20.33){\makebox(0,0)[cc]{$_q$}}
\put(65.67,9.67){\makebox(0,0)[cc]{$_p$}}
\put(18.00,10.33){\makebox(0,0)[cc]{$_p$}}
\put(18.00,19.00){\makebox(0,0)[cc]{$_q$}}
\put(27.00,19.00){\makebox(0,0)[cc]{$_k$}}
\put(27.33,10.33){\makebox(0,0)[cc]{$_{-p-q-k}$}}
\special{em:linewidth 1.2pt}
\linethickness{1.2pt}
\emline{125.00}{19.00}{1}{133.00}{19.00}{2}
\emline{133.00}{19.00}{3}{133.00}{11.00}{4}
\emline{133.00}{11.00}{5}{125.00}{11.00}{6}
\emline{125.00}{11.00}{7}{125.00}{19.00}{8}
\put(57.00,15.00){\makebox(0,0)[cc]{$+$}}
\end{picture}
\vspace{-20mm}
\begin{equation}
\label{MG4}
\end{equation}
Equation (\ref{MG4})
states that 1PI   contributions into the
4-point Green's function $\Gamma_4$ are presented by the sum of the bare
4-point vertex,  horizontal and vertical vertex functions
and diagrams that are 2PR in both $s$- and $t$-channels.

Equations (\ref{MG3})-(\ref{MG4})
together with the  Schwinger-Dyson equation for propagator
(\ref{SDD})
form the closed system of non-linear integral equations
for the functions $D(p)$, $\Gamma _3(p,q)$,
$\Gamma _4 (p,q,k)$,
$ H(p,q,k)$ and
$ V(p,q,k)$.
Just these equation we call the planar parquet equations
for the basic functions.
Note that the planar parquet equations are  different from
the usual parquet equations \cite{TM} since
in the planar approximation so-called $u$-channel terms do not contribute.

\section{Weighted Number of Planar Parquet Diagrams for Quartic Interaction}

In this section we examine the
weighed number of planar parquet diagrams
in the matrix theory with quartic interaction. This means that the
dependence on momenta  variables in (\ref{SDD})-(\ref{MG4})
must be neglected. In this case equations   (\ref{SDD})-(\ref{MG4})
are reduced
to the following system of algebraic equations
\begin{equation}
\label{4D}
D=1-2 g D^2 - g D^4 \Gamma_4.
\end{equation}
\begin{equation}
\label{4G4}
\Gamma_4 =-g +H+V,
\end{equation}
\begin{equation}
\label{4H}
H=-g D^2 \Gamma_4 +V D^2 \Gamma_4,
\end{equation}
\begin{equation}
\label{4V}
V=-g D^2 \Gamma_4 +H D^2 \Gamma_4.
\end{equation}
The graphical representation of equations (\ref{4D})-(\ref{4V}) is
given on Figure 1.

\vspace{5mm}

\unitlength=1.00mm
\special{em:linewidth 0.4pt}
\linethickness{0.4pt}
\begin{picture}(150.00,66.30)
\special{em:linewidth 0.4pt}
\linethickness{0.4pt}
\put(75.00,0.00){\makebox(0,0)[cc]{Figure 1.Graphical representation of
equations (\ref{4D})-(\ref{4H}). }} \put(124.00,17.00){\line(0,-1){1.67}}
\put(35.00,15.00){\rule{12.00\unitlength}{4.00\unitlength}}
\put(80.00,17.00){\circle*{5.20}}
\put(110.00,11.00){\rule{4.00\unitlength}{12.00\unitlength}}
\put(124.00,17.00){\circle*{5.20}}
\put(96.00,17.00){\makebox(0,0)[cc]{$+$}}
\put(58.00,17.00){\makebox(0,0)[cc]{$=$}}
\put(40.00,40.00){\circle*{5.20}}
\put(85.00,38.00){\rule{12.00\unitlength}{4.00\unitlength}}
\put(120.00,34.00){\rule{4.00\unitlength}{12.00\unitlength}}
\put(109.00,40.00){\makebox(0,0)[cc]{$+$}}
\put(75.00,40.00){\makebox(0,0)[cc]{$+$}}
\put(53.00,40.00){\makebox(0,0)[cc]{$=$}}
\put(75.33,64.00){\circle{4.00}}
\put(102.00,60.00){\circle{4.00}}
\put(75.33,64.00){\circle{4.20}}
\put(102.00,60.00){\circle{4.20}}
\put(75.33,64.00){\circle{4.40}}
\put(102.00,60.00){\circle{4.40}}
\put(75.33,64.00){\circle{4.60}}
\put(102.00,60.00){\circle{4.60}}
\put(114.00,62.00){\makebox(0,0)[cc]{$+$}}
\put(88.33,62.00){\makebox(0,0)[cc]{$+$}}
\put(62.00,62.00){\makebox(0,0)[cc]{$+$}}
\put(38.00,62.00){\makebox(0,0)[cc]{$=$}}
\put(126.00,44.00){\line(-1,0){8.00}}
\put(118.00,36.00){\line(1,0){8.00}}
\put(95.00,36.00){\line(0,1){8.00}}
\put(87.00,44.00){\line(0,-1){8.00}}
\put(37.00,43.00){\line(1,-1){6.00}}
\put(43.00,43.00){\line(-1,-1){6.00}}
\put(108.00,13.00){\line(1,0){3.00}}
\put(111.00,21.00){\line(-1,0){3.00}}
\put(124.00,18.67){\line(1,0){4.00}}
\put(124.00,15.33){\line(1,0){4.00}}
\put(81.33,18.67){\line(1,0){2.67}}
\put(84.00,15.33){\line(-1,0){2.67}}
\put(37.00,21.00){\line(0,-1){8.00}}
\put(45.00,13.00){\line(0,1){8.00}}
\put(94.00,62.00){\line(1,0){8.00}}
\put(56.00,62.00){\line(-1,0){12.00}}
\put(68.00,62.00){\line(1,0){7.33}}
\put(71.00,17.00){\line(-3,1){4.00}}
\put(71.00,17.00){\line(-3,-1){4.00}}
\put(62.67,41.67){\line(4,-3){4.67}}
\put(138.00,62.00){\circle*{5.20}}
\put(67.33,41.67){\line(-4,-3){4.67}}
\put(120.00,62.00){\line(1,0){8.00}}
\special{em:linewidth 1.2pt}
\linethickness{1.2pt}
\emline{71.00}{17.00}{1}{80.00}{18.67}{2}
\emline{80.00}{15.33}{3}{71.00}{17.00}{4}
\emline{112.00}{21.00}{5}{124.00}{18.67}{6}
\emline{124.00}{15.33}{7}{112.33}{13.00}{8}
\emline{127.00}{62.00}{9}{150.00}{62.00}{10}
\emline{127.00}{62.00}{11}{138.00}{63.67}{12}
\emline{138.00}{60.33}{13}{128.00}{62.00}{14}
\emline{109.00}{62.00}{15}{102.00}{62.00}{16}
\emline{83.00}{62.00}{17}{75.33}{62.00}{18}
\emline{30.00}{62.00}{19}{17.00}{62.00}{20}
\end{picture}

\vspace{5mm}

Four equations (\ref{4D})-(\ref{4V}) contain four unknown functions
$D(g),~\Gamma (g),~H(g),~V(g)$ and can be solved explicitly.
Excluding $\Gamma _4$, $H$ and $V$   from
(\ref{4D})-(\ref{4V})  we get
\begin{equation}
\label{4eq1}
g^3 D^6+g^2 D^5
+5 g^2 D^4+ 5 g D^3 + (1-5 g) D^2 - 2 D +1=0.
\end{equation}
Below on the second  line of  Tables 1 we present
numerical solutions of (\ref{4eq1})
for different values of the coupling constant
$g$.  Corresponding $\Gamma_4$ is presented on the fourth line
of this Table.
\vspace{10mm}

Table 1.

\begin{tabular}{llllllll}
\hline
$g$ & $0.001$ & $0.01$ & $0.1$ & $1$ & $10$ & $100$ & $1000$
\\
\hline
$D$
& $0.9980$
& $0.9808$
& $0.8576$
& $0.5161$
& $0.2129$
& $0.07372$
& $0.02400$
\\
$D^{(pl)}$
& $0.9980$
& $0.9808$
& $0.8576$
& $0.5161$
& $0.2130$
& $0.07374$
& $0.02401$

\\
\hline
$\Gamma _4$
& $-9.980 \cdot 10^{-4}$
& $-9.813 \cdot  10^{-3}$
& $-0.08786$
& $-0.6896$
& $-5.823$
& $-54.38$
& $-531.3$

\\
$\Gamma _4^{(pl)}$
& $-9.980 \cdot 10^{-4}$
& $-9.813 \cdot 10^{-3}$
& $-0.08786$
& $-0.6900$
& $-5.835$
& $-54.55$
& $-533.1$
\\
\hline
\end{tabular}
\vspace{5mm}

It is instructive to compare these results
with the exact weighted numbers of the planar diagrams.
These numbers are solutions of the corresponding  planar
theory.
The zero-dimensional planar Green's functions $\Pi _n$ for the quartic
interaction satisfy the Schwinger-Dyson equations
\begin{equation}
\label{0SDE}
\Pi_n=-g\Pi_{n+2} +\sum_{m=2}^{n}\Pi_{m-2} \Pi_{n-m},
\end{equation}
that are zero-dimensional analogs of (\ref{PSDE}).
These equations admit the following solution
\begin {equation}
\label {EqPi4}
\Pi _4=g^{-1}(1-D^{(pl)}),
\end   {equation}
where the planar propagator $D^{(pl)} \equiv
\Pi_2 $ satisfies the  equation
\begin{equation}
\label{eqDpl}
27 g^2 D^{(pl)~2} +(1+18g) D^{(pl)}  -1-16g=0.
\end{equation}
The numerical values for the planar propagator $D^{(pl)}$ and
the four-point vertex
$\Gamma _4^ {(pl)}$ $=$ $(D^{(pl)})^{-4}[\Pi _4-2(D^{(pl)})^2]$
are also presented on Table 1.

We see that numerical values of  Green's functions
in the planar parquet approximation and in the planar approximation
for equal values of the coupling constant are strikingly closed, namely the
difference is less then 0.1 percent for all range of $g$.

It is also instructive to compare an asymptotic behaviour of Green's functions in the planar
parquet approximation with the planar approximation.
  From  equation (\ref{4eq1})  we have the following
asymptotic behavior of $D(g)$
in the strong coupling regime
\begin{equation}
\label{4aD1}
D(g) \sim \alpha_{par} g^{-1/2},
\end{equation}
where $\alpha_{par}$ is the solution of an equation
\begin{equation}
\label{4aD2}
\alpha_{par} ^6+
5 \alpha_{par} ^4-
5 \alpha_{par}^2 +1=0.
\end{equation}
The numerical solution of equation (\ref{4aD2}) is
\begin{equation}
\label{4aD3}
\alpha_{par}=0.76948~.
\end{equation}
The asymptotic behavior of the  planar propagator
$D^{(pl)}$ follows from equation (\ref{eqDpl})
\begin{equation}
\label{4aplD1}
D^{(pl)}(g) \sim \alpha_{pl} g^{-1/2},
\end{equation}
where
\begin{equation}
\label{4aplD3}
\alpha_{pl}=\frac{4\sqrt{3}}{9}=0.76980~.
\end{equation}
The agreement of (\ref{4aD3}) and (\ref{4aplD3}) is more then
susceptible.

We also compare propagator and the four-vertex function
in the parquet approximation with the exact results for negative
values of the coupling constant (see Table 2).

\vspace{5mm}

Table 2.

\begin{tabular}{lllllllll}
\hline
$g$ & $-0.001$ & $-0.01$ & $-0.05$ & $-0.08$ &$-0.0833$&
$-0.0834$ & $-0.0864$
& $-0.0865$
\\
\hline
$D$ & $1.0020$ & $1.0210$ & $1.1332$ & $1.2959$ &  $~$ & $~$ &
$1.3889$ & $~~-$
\\
$D^{(pl)}$ & $1.0020$ & $1.0210$ & $1.1332$ & $1.2963$ &
$1.333$ & $~~-$ & $~$ & $~$

\\
\hline
$\Gamma _4$ & $0.0010$
& $0.0102$
& $0.05806$
& $0.1207$
&$~$
&$~$
& $0.1728$
& $~~-$

\\
$\Gamma _4^{(pl)}$ & $0.0010$
& $0.0102$
& $0.05806$
& $0.1214$
& $0.1403$
& $~~-$
& $~$
& $~$
\\
\hline
\end{tabular}
\vspace{10mm}

Remarkably that the solution of the parquet equation has a phase
transition for $g=g_0$, $-0.0865<g_0<-0.0864$  that is related with the
root branch point of the solution.  In the planar approximation  the
phase transition point is caused by the square root branch point and
it is equal to $g_0=-1/12=-0.0833(3)$.
\section{Weighted Numbers of Planar Parquet Diagrams for
Cubic Interaction}

In this section we
consider the weighed numbers
of planar parquet diagrams for the matrix model
with the cubic interaction.
For this case  planar parquet equations  (\ref{SDD})-(\ref{MG4})
reduce to the form
\begin{equation}
\label{D}
D=1+  \lambda  D^3 \Gamma_3,
\end{equation}
\begin{equation}
\label{parG3}
\Gamma_3 =  \lambda +D^2 V \Gamma_4 + D^3 \Gamma _3^3
\end{equation}
\begin{equation}
\label{parG}
\Gamma _4 = H + V +D^4 \Gamma_3^4,
\end{equation}
\begin{equation}
\label{parH}
H=D^2 \Gamma_4 V+
D^3 \Gamma_3^2 V+
D^3 \Gamma_4 \Gamma_3^2,
\end{equation}
\begin{equation}
\label{parV}
V=D^2 \Gamma_4 H+
D^3 \Gamma_3^2 H+
D^3 \Gamma_4 \Gamma_3^2.
\end{equation}
A graphical representation of these equations
is given on Figure 2.

\vspace{5mm}

\unitlength=1.00mm
\special{em:linewidth 0.4pt}
\linethickness{0.4pt}
\begin{picture}(137.00,90.60)
\special{em:linewidth 0.4pt}
\linethickness{0.4pt}
\put(68.00,17.00){\line(0,-1){1.67}}
\put(25.00,15.00){\rule{12.00\unitlength}{4.00\unitlength}}
\put(54.00,11.00){\rule{4.00\unitlength}{12.00\unitlength}}
\put(68.00,17.00){\circle*{5.20}}
\put(33.00,40.00){\circle*{5.20}}
\put(55.00,38.00){\rule{12.00\unitlength}{4.00\unitlength}}
\put(90.00,34.00){\rule{4.00\unitlength}{12.00\unitlength}}
\put(79.00,40.00){\makebox(0,0)[cc]{$+$}}
\put(96.00,44.00){\line(-1,0){8.00}}
\put(88.00,36.00){\line(1,0){8.00}}
\put(65.00,36.00){\line(0,1){8.00}}
\put(57.00,44.00){\line(0,-1){8.00}}
\put(30.00,43.00){\line(1,-1){6.00}}
\put(36.00,43.00){\line(-1,-1){6.00}}
\put(52.00,13.00){\line(1,0){3.00}}
\put(55.00,21.00){\line(-1,0){3.00}}
\put(68.00,18.67){\line(1,0){4.00}}
\put(68.00,15.33){\line(1,0){4.00}}
\put(27.00,21.00){\line(0,-1){8.00}}
\put(35.00,13.00){\line(0,1){8.00}}
\put(33.00,65.00){\circle*{5.20}}
\put(28.00,65.00){\line(1,0){5.00}}
\put(36.67,68.67){\line(-1,-1){3.67}}
\put(33.00,65.00){\line(1,-1){3.67}}
\put(125.00,70.00){\circle*{3.33}}
\put(125.00,60.00){\circle*{3.33}}
\put(88.00,12.00){\circle*{3.33}}
\put(88.00,22.00){\circle*{3.33}}
\put(100.00,17.00){\circle*{5.20}}
\put(88.00,22.00){\line(-1,0){3.67}}
\put(84.33,12.00){\line(1,0){3.67}}
\put(100.00,18.67){\line(1,0){4.33}}
\put(100.00,15.33){\line(1,0){4.33}}
\put(118.00,11.00){\rule{4.00\unitlength}{12.00\unitlength}}
\put(120.00,21.00){\line(-1,0){4.00}}
\put(120.00,13.00){\line(-1,0){4.00}}
\put(133.00,13.00){\circle*{3.33}}
\put(133.00,21.00){\circle*{3.33}}
\put(110.00,17.00){\makebox(0,0)[cc]{$+$}}
\put(78.00,17.00){\makebox(0,0)[cc]{$+$}}
\put(45.00,17.00){\makebox(0,0)[cc]{$=$}}
\put(45.00,40.00){\makebox(0,0)[cc]{$=$}}
\put(115.00,36.00){\circle*{3.33}}
\put(115.00,44.00){\circle*{3.33}}
\put(123.00,44.00){\circle*{3.33}}
\put(123.00,36.00){\circle*{3.33}}
\put(123.00,36.00){\line(1,0){3.67}}
\put(123.00,44.00){\line(1,0){3.67}}
\put(115.00,44.00){\line(-1,0){3.67}}
\put(111.33,36.00){\line(1,0){3.67}}
\put(125.00,70.00){\line(1,0){3.67}}
\put(125.00,60.00){\line(1,0){3.67}}
\put(104.00,40.00){\makebox(0,0)[cc]{$+$}}
\put(115.00,65.00){\line(-1,0){4.00}}
\put(65.00,88.00){\line(1,0){12.00}}
\put(112.00,88.00){\circle*{5.20}}
\put(100.00,88.00){\line(-1,0){10.00}}
\put(83.67,88.00){\makebox(0,0)[cc]{$+$}}
\put(58.00,88.00){\makebox(0,0)[cc]{$=$}}
\put(103.00,65.00){\makebox(0,0)[cc]{$+$}}
\put(44.33,65.00){\makebox(0,0)[cc]{$=$}}
\put(57.00,65.00){\line(6,5){4.00}}
\put(57.00,65.00){\line(6,-5){4.00}}
\put(53.00,65.00){\line(1,0){4.00}}
\put(112.00,90.00){\line(0,-1){3.67}}
\put(68.00,18.67){\line(0,-1){3.33}}
\put(67.00,65.00){\makebox(0,0)[cc]{$+$}}
\put(133.00,13.00){\line(1,0){4.00}}
\put(133.00,21.00){\line(1,0){4.00}}
\put(116.00,65.00){\circle*{3.33}}
\put(80.00,65.00){\circle*{5.20}}
\put(91.00,59.00){\rule{4.00\unitlength}{12.00\unitlength}}
\put(93.00,69.00){\line(1,0){4.00}}
\put(93.00,61.00){\line(1,0){4.00}}
\put(80.00,65.00){\line(-1,0){5.67}}
\put(75.00,0.00){\makebox(0,0)[cc]{Figure 2. Graphical representation of
equations (\ref{D})-(\ref{parH}).}}
 \special{em:linewidth 1.2pt}
\linethickness{1.2pt}
\emline{56.00}{21.00}{1}{68.00}{18.67}{2}
\emline{68.00}{18.67}{3}{68.00}{15.33}{4}
\emline{68.00}{15.33}{5}{56.00}{13.00}{6}
\emline{88.00}{12.00}{7}{88.00}{22.00}{8}
\emline{88.00}{22.00}{9}{100.00}{18.67}{10}
\emline{100.00}{15.33}{11}{88.00}{12.00}{12}
\emline{120.00}{21.00}{13}{133.00}{21.00}{14}
\emline{133.00}{21.00}{15}{133.00}{13.00}{16}
\emline{133.00}{13.00}{17}{120.00}{13.00}{18}
\emline{123.00}{44.00}{19}{123.00}{36.00}{20}
\emline{123.00}{36.00}{21}{115.00}{36.00}{22}
\emline{115.00}{36.00}{23}{115.00}{44.00}{24}
\emline{115.00}{44.00}{25}{123.00}{44.00}{26}
\emline{115.00}{65.00}{27}{125.00}{60.00}{28}
\emline{125.00}{60.00}{29}{125.00}{70.00}{30}
\emline{125.00}{70.00}{31}{115.00}{65.00}{32}
\emline{100.00}{88.00}{33}{112.00}{89.67}{34}
\emline{112.00}{86.33}{35}{101.00}{88.00}{36}
\emline{51.00}{88.00}{37}{35.00}{88.00}{38}
\emline{80.00}{66.67}{39}{93.00}{69.00}{40}
\emline{93.00}{69.00}{41}{93.00}{61.00}{42}
\emline{93.00}{61.00}{43}{80.00}{63.33}{44}
\emline{112.00}{88.00}{45}{123.00}{88.00}{46}

\end{picture}

\vspace{5mm}

Note that
using equations (\ref{parG})-(\ref{parV})
one can rewrite the equation for $\Gamma_3$ in the form
\begin{equation}
\label{parG3a}
\Gamma_3= \lambda   + \lambda   D^2 \Gamma_4
+  \lambda  D^3 \Gamma_3^2 ,
\end{equation}
or graphically

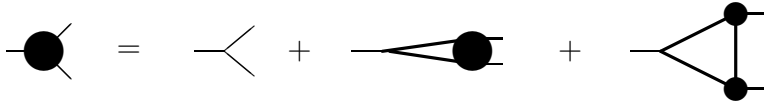
\unitlength=1.00mm
\special{em:linewidth 0.4pt}
\linethickness{0.4pt}
\begin{picture}(128.67,21.66)
\special{em:linewidth 0.4pt}
\linethickness{0.4pt}
\put(33.00,15.00){\circle*{5.20}}
\put(28.00,15.00){\line(1,0){5.00}}
\put(36.67,18.67){\line(-1,-1){3.67}}
\put(33.00,15.00){\line(1,-1){3.67}}
\put(90.00,15.00){\circle*{5.20}}
\put(94.00,16.67){\line(-1,0){4.00}}
\put(90.00,16.67){\line(0,-1){3.33}}
\put(90.00,13.33){\line(1,0){4.00}}
\put(125.00,20.00){\circle*{3.33}}
\put(125.00,10.00){\circle*{3.33}}
\put(125.00,20.00){\line(1,0){3.67}}
\put(125.00,10.00){\line(1,0){3.67}}
\put(78.00,15.00){\line(-1,0){4.00}}
\put(115.00,15.00){\line(-1,0){4.00}}
\put(103.00,15.00){\makebox(0,0)[cc]{$+$}}
\put(44.33,15.00){\makebox(0,0)[cc]{$=$}}
\put(57.00,15.00){\line(6,5){4.00}}
\put(57.00,15.00){\line(6,-5){4.00}}
\put(53.00,15.00){\line(1,0){4.00}}
\put(90.00,16.67){\line(0,-1){3.33}}
\put(67.00,15.00){\makebox(0,0)[cc]{$+$}}
\special{em:linewidth 1.2pt}
\linethickness{1.2pt}
\emline{115.00}{15.00}{1}{125.00}{10.00}{2}
\emline{125.00}{10.00}{3}{125.00}{20.00}{4}
\emline{125.00}{20.00}{5}{115.00}{15.00}{6}
\emline{78.00}{15.00}{7}{90.00}{16.67}{8}
\emline{90.00}{13.33}{9}{79.00}{15.00}{10}
\end{picture}
\vspace{-20mm}
$$~$$
Equation  (\ref{parG3a})
is just the Schwinger-Dyson equation for the 1PI three-point
Green's function $\Gamma _3$.

Excluding $\Gamma_3$, $\Gamma _4$, $H$ and $V$
from equations  (\ref{parG})-(\ref{parV})  we get the equation for $D$
\begin{equation}
\label{eqD}
2  \lambda  ^6 D^9-3  \lambda   ^4 D^6(D-1)+
 \lambda  ^2 D^4 (D-1)^2 -(D-1)^5=0.
\end{equation}

Let us compare  the solution of the planar parquet equations
(\ref{parG})-(\ref{parG3}) with the solution of the planar theory.
The  planar 2-point and 1PI 3-point Green's function are
given by formulas \cite{BIPZ}:
\begin{equation}
\label{plD}
D^{(pl)}=\frac{1-3\tau}{(1-2\tau )^2},~~~
\Gamma_3^{(pl)}=  \lambda  \frac{(1-2\tau)^2(1-4 \tau)}{(1-3 \tau)^3},
\end{equation}
where $\tau$ is the subject of the relation
\begin{equation}
\label{tau}
 \lambda ^2=\tau (1-2 \tau)^2.
\end{equation}

The results of  numerical solutions of planar parquet equations
(\ref{parG})-(\ref{parG3})  as well as equations
(\ref{plD})-(\ref{tau}) for
planar Green's functions
are presented on Tables 3 and 4.

\vspace{5mm}

Table 3. Perturbative solution.

\begin{tabular}{llllllllll}
\hline
$ \lambda $ & $0.001$ & $0.01$ & $0.1$ & $0.2$ & $0.27$
 & $0.272$ & $0.273$
&$0.279$ & $0.280$
\\
\hline
$D$ & $1.00$ & $1.00$ & $1.0104$ & $1.0486$ &
$1.1197$ & $~$ &  $~$ & $1.1458$ & $~~ - $
\\
$D^{(pl)}$
& $1.00$
& $1.00$
& $1.0104$
& $1.0486$
& $1.1200$
& $1.1245$
& $~~-$
& $~$ &  $~$

\\
\hline
$\Gamma _3$ & $ 0.0010$ & $0.010$ & $0.10106$
& $0.21084$  &
$0.31577$ & $~$ &  $~$ & $0.34720$ & $~~-$

\\
$\Gamma _3^{(pl)}$
& $0.0010$
& $0.010$
& $0.10106$
& $0.21084$
& $0.31644$
& $0.32202$
& $~~-$
& $~$ &  $~$
\\
\hline
\end{tabular}
\vspace{5mm}

Table 4.  Nonperturbative solution.

\begin{tabular}{llllllll}
\hline
$ \lambda $ & $0.001$ & $0.01$ & $0.1$ & $1$ & $10$ & $100$ & $1000$
\\
\hline
$D$
& $-26783$
& $-864$
& $-30.78$
& $-1.96$
& $-0.286$
& $-0.0565$
& $-0.0119$
\\
$D^{(pl)}$
& $-2.51 \cdot  10^5$
& $-2640$
& $-39.6$
& $-2.00$
& $-0.287$
& $-0.0566$
& $-0.0120$

\\
\hline
$\Gamma _3$
& $1.39 \cdot 10^{-6}$
& $1.34 \cdot 10^{-4}$
& $0.0109$
& $0.0.391$
& $5.48$
& $58.6$
& $594$

\\
$\Gamma _3^{(pl)}$
& $1.58 \cdot 10^{-8}$
& $1.43 \cdot 10^{-5}$
& $0.00653$
& $0.375$
& $5.42$
& $58.2$
& $590$
\\
\hline
\end{tabular}
\vspace{5mm}

Table 3 corresponds to a perturbative solution.
When $ \lambda =0$ we have
$D=1$ and $\Gamma_3=0$ that is related to the free theory. There is a phase
transition point
$ \lambda =\lambda_0$ where $0.279< \lambda_0<0.280 $ for the planar
parquet approximation and $\lambda_0 =\sqrt{2} / ( 3\sqrt{3} )$ $=$
$0.2722 $ for the planar approximation.
Table 4 corresponds to another branch that is a
nonperturbative solution.

  From (\ref{eqD}) we find
the following asymptotic behavior for $D$ in the strong coupling regime
\begin{equation}
\label{asD}
D(g) \sim c_{par}   \lambda  ^{-2/3},
\end{equation}
where $c_{par} $ is satisfied  the equation
\begin{equation}
\label{c}
2 c_{par}^9 +3 c_{par}^6 +1=0.
\end{equation}
The numerical solution of equation (\ref{c}) is
\begin{equation}
\label{cc}
c_{par}=-1.18823~.
\end{equation}
An asymptotic behavior of  the planar propagator is found from (\ref{plD}),
(\ref{tau}) and has the form:
\begin{equation}
\label{plasD}
D^{(pl)}( \lambda )
\sim c_{pl}  \lambda ^{-2/3}
\end{equation}
with
\begin{equation}
\label{plcc}
c_{pl}=-\frac{3(4)^{1/3}}{4}=-1.19055~.
\end{equation}

\section{Conclusion}

In conclusion, in this paper we found that the planar parquet
 approximation gives a unexpectedly good agreement with the known exact
 results for large $N$ matrix models.
The parquet approximation in quantum field theory
is one of few known approximations
which admits a non-perturbative closed set of equations
for finite Green's functions.

The results of this paper mean that the class of parquet diagrams
is rather comprehensive among of all
planar diagrams of matrix theories. This makes
reasonable an investigation of planar parquet approximation for QCD
that will be the subject of our further work.


$$~$$
{\bf ACKNOWLEDGMENT}

\vspace{5mm}

Both authors are supported by RFFR
grant 96-01-00608.
We are grate\-ful to
G.E.Aru\-tyu\-nov, P.B.Med\-ve\-dev and I.Vo\-lo\-vich for use\-ful
dis\-cussions.
$$~$$


\end{document}